\documentclass{amsart}
\usepackage{amssymb,amsmath,amsthm,mathrsfs,multirow,enumerate,mathtools}
\usepackage{braket}
\usepackage{graphicx} 
\usepackage[all]{xy}

\usepackage{xcolor}
\numberwithin{equation}{section} 

\theoremstyle{plain}
\newtheorem{thm}{Theorem}[section]

\newtheorem{lemma}[thm]{Lemma}

\theoremstyle{definition}

\newtheorem{ex}{Example}

\newtheorem*{rmk}{Remark}

\newtheorem{quest}{Question}

\newcommand{\bbm}{\begin{bmatrix}}
\newcommand{\ebm}{\end{bmatrix}}

\begin{document}

\title[CS matrices from orthogonal spaces]{Compressed sensing matrices from orthogonal spaces over finite fields of odd characteristic}

\author{Kanittakorn Moonchaisook}
\address{Kanittakorn Moonchaisook, KMUTT-Fixed Point Theory and Applications Research Group, and 
Department of Mathematics, Faculty of Science, King Mongkut’s University of Technology Thonburi (KMUTT), 
126 Pracha-Uthit Road, Bang Mod, Thrung Khru, Bangkok 10140, Thailand} 
\email{kanittareal@gmail.com}

\author{Poom Kumam}
\address{Poom Kumam,  KMUTT-Fixed Point Theory and Applications Research Group, and 
Department of Mathematics, Faculty of Science, King Mongkut’s University of Technology Thonburi (KMUTT), 
126 Pracha-Uthit Road, Bang Mod, Thrung Khru, Bangkok 10140, Thailand}
\email{\tt poom.kum@kmutt.ac.th}

 \author{Songpon Sriwongsa$^*$}
\thanks{*Corresponding Author}
\address{Songpon Sriwongsa, KMUTT-Fixed Point Theory and Applications Research Group, and 
Department of Mathematics, Faculty of Science, King Mongkut’s University of Technology Thonburi (KMUTT), 
126 Pracha-Uthit Road, Bang Mod, Thrung Khru, Bangkok 10140, Thailand}
\email{\tt  songponsriwongsa@gmail.com, songpon.sri@kmutt.ac.th}



\begin{abstract}
In this paper, we construct deterministic matrices from subspaces of orthogonal spaces over finite fields of odd characteristic and investigate their applicability to compressed sensing. The construction is based on incidence relations among three types of subspaces, yielding families of matrices with explicitly computable dimensions and coherence. Using coherence-based estimates, we establish sufficient conditions under which these matrices satisfy the Restricted Isometry Property for prescribed sparsity levels. We also provide numerical comparisons with DeVore's deterministic construction to illustrate the trade-off between the number of measurements, coherence, and sparse recovery guarantees.
\end{abstract}

\maketitle

\section{Introduction}

Compressed Sensing (CS) is a revolutionary paradigm in signal processing and information theory that fundamentally challenges the Nyquist-Shannon sampling theorem. Rather than requiring measurements at a rate proportional to a signal's bandwidth, CS theory establishes that a sparse signal, one with few non-zero elements in a known basis, can be faithfully recovered from a significantly smaller number of linear measurements. This efficiency directly translates into a reduction in the time, cost, and energy required for signal acquisition, driving its adoption across diverse applications such as Magnetic Resonance Imaging (MRI), analog-to-digital conversion, and data compression.

The process of CS begins with a discrete signal $\mathbf{x} \in \mathbb{R}^N$. Instead of acquiring all $N$ components, we capture a smaller set of $M$ linear measurements. These projections are encapsulated in an $M \times N$ matrix $\mathbf{\Phi}$, known as the \textit{CS matrix}. The resulting vector $\mathbf{y} \in \mathbb{R}^M$, defined by the relationship $\mathbf{y} = \mathbf{\Phi x}$, is termed the information vector or measurement vector. The effectiveness of this scheme relies on the signal's inherent structure, specifically its sparsity: if $\mathbf{x}$ has at most $k$ non-zero entries, it is identified as \textit{$k$-sparse}. Given this framework, the central question for signal recovery immediately emerges:
\begin{quest}
Can we reconstruct the original signal $\mathbf{x}$ from the compressed measurement vector $\mathbf{y} = \mathbf{\Phi x}$?
\end{quest}

 Candes et al. \cite{Can2} and Donoho \cite{D06} proposed to use few measurements to obtain the information of the original signal by sufficiently using the feature of a sparse signal, particularly, the sparse solution of the linear equation seeks 
 \[
 y=\Phi x : \displaystyle\min_{x\in\mathbb{R}^{N}}||x||_0.
 \] 

Later, Candes \cite{Can} introduced a mathematical condition, namely the {\it Restricted Isometry Property} (RIP) as follows. Let $\Phi$ be an $M\times N$ matrix. We say that $\Phi$ satisfies the {\it RIP} of order $k$ with constant $\delta_k \in (0, 1)$ if for all $k$-sparse signal $x\in \mathbb{R}^N$, the following inequality holds:
\[ 
(1-\delta_k)||x||^2_2\leq ||\Phi x||^2_2\leq (1+\delta_k)||x||^2_2, 
\]
the minimum nonnegative constant $\delta_k$ is called the {\it RIP constant of order} $k$. Suppose that
\begin{equation}
  \delta_{2k} < \sqrt{2} - 1.
  \label{eq:rip-condition}
\end{equation}
Then the following convex optimization problem, known as {\it basis pursuit},
\begin{equation}
  \min_{\xi} \lVert \xi \rVert_1
  \quad \text{such that} \quad y = \Phi \xi,
  \label{eq:basis-pursuit}
\end{equation}
where the $\ell_1$ norm $\lVert \xi \rVert_1$ of $\xi$ is defined as the sum of the absolute values of the entries of $\xi$, has $x$ as its unique solution. Thus, we would like to construct CS matrices $\Phi$ that have both small row dimension $M$, so that they provide as economic measurements of $x$ as possible, while at the same time having small RIP constant $\delta_{2k} < \sqrt{2} - 1$, so that they can be used for the recovery of $x$ by solving \eqref{eq:basis-pursuit}.

 Let $\alpha_1, \alpha_2, ..., \alpha_N$ be column vectors of $\Phi$ . Let $\mu(\Phi)$ denote the coherence of a matrix $\Phi$, 
\[ \mu(\Phi)=\max_{1\leq i,j\leq N}\frac{|\langle \alpha_i,\alpha_j \rangle|}{||\alpha_i||_2||\alpha_j||_2} \]
where $\langle \alpha_i,\alpha_j \rangle$ is the inner product between two columns  $\alpha_i,\alpha_j$ of $\Phi$.
Then the following property holds.

\begin{lemma}\label{Bglem}\cite{Bour}  If unit vectors $\alpha_1, \alpha_2, ..., \alpha_N$ are the columns of a matrix $\Phi$ and a coherence $\mu=\mu(\Phi),$ then $\Phi$ satisfies the RIP of order $k$ with $\delta_k\leq (k-1)\mu$ for all $k<1+\frac{1}{\mu}$.
\end{lemma}

The design of the Compressed Sensing (CS) matrix is fundamental to the entire CS framework. While random sensing matrices perform well both theoretically and empirically, their main drawback is their limited effectiveness, in fact, they tend to satisfy RIP only for relatively low sparsity levels $k$. To overcome this constraint, DeVore \cite{DeV} pioneered the use of deterministic sensing matrices. Since low coherence between the columns of the matrix is a strong indicator of the RIP, coherence has become a central consideration in crafting these deterministic constructions. The method in this reference is a notable example, successfully yielding matrices that satisfy the RIP for significantly higher orders of sparsity. The specific construction is outlined below.

Let $p$ be a prime number and let $\mathbb{F}_{p^d}$ be a finite field of order $p^d$.  Let $\mathcal{P}_r$ be the set of polynomials over $\mathbb{F}_{p^d}$ with degree
$\leq r$. Note that $|\mathcal{P}_r| = p^{d(r + 1)}$. Any $f\in \mathcal{P}_r$ can be viewed as a map on $\mathbb{F}_{p^d}$ and expressed in a binary vector form, noted by $V_f$. The form of $V_f$ is 
\[ f_{i,j} = \begin{cases}
	1 & \text{if} \ f(i)=j,\\
	0 & \text{else.}
\end{cases} \] 
There are exactly $p^d$ ones in $V_f, f \in \mathcal{P}_r,$ constitute a $p^{2d}\times p^{d(r+1)}$ matrix $\Phi_0$. There is no
more than $r$ inner product of any two distinct columns in $\Phi_0$.
Since the difference of two distinct polynomials is a nonzero polynomial belonging to $\mathcal{P}_r,$ it has no more than $r$ roots. 

\begin{thm}\label{DevThm}\cite{DeV} Let $\Phi_0$ be the $p^{2d}\times p^{d(r+1)}$ matrix with column $V_f, f\in \mathcal{P}_r.$ Then the matrix $\Phi=\frac{1}{\sqrt{p^d}}\Phi_0$ satisfies RIP with $\delta_k\leq \frac{(k-1)r}{p^d}$ for any $k < \frac{p^d}{r+1}$.
\end{thm}

A significant alternative approach to CS matrix design involves exploiting the inherent structure of finite geometry over finite fields. 
In 2016, Guo and Zhang \cite{CC} provided two constructions of the CS matrix based on symplectic geometry and singular symplectic geometry over finite fields. In fact, the subspaces of type $(m, s)$ in symplectic spaces and the subspaces of type $(m, s, k)$ in singular symplectic spaces were used. For similar constructions from pseudo-symplectic spaces and unitary geometry over finite fields, see \cite{GFX, TLPY}. Motivated by these works, in this paper, we construct CS matrices from orthogonal spaces over finite fields of odd characteristic. 
We first recall the theory of these spaces as follows (see \cite{Wan}).

 Let $S$ be $n \times n$ nonsingular symmetric matrix over a finite field $\mathbb{F}_q$ of order odd prime power $q$. An $n\times n$ matrix $T$ over $\mathbb{F}_q$ is said to be {\it orthogonal with respect to} $S$, if
\[
TST^T = S.
\]
Clearly, $n\times n$ orthogonal matrices with respect to a nonsingular symmetric matrix $S$ are nonsingular and they form a group with respect to matrix multiplication, called the {\it orthogonal group of degree $n$ with respect to $S$ over} $\mathbb{F}_q$ and denoted by $O_n(\mathbb{F}_q,S)$. If $S_1$ and $S_2$ are two cogredient $n\times n$ nonsingular symmetric matrices over $\mathbb{F}_q$, then as in the symplectic case we can prove that $O_n(\mathbb{F}_q,S_1)$ and $O_n(\mathbb{F}_q,S_2)$ are isomorphic. It is well-known that the matrix $S$ is cogredient to 
\begin{align*}
    S_{2\nu + \delta, \Delta} = 
    \begin{pmatrix}
        & I_\nu &  \\
    I_\nu &     &  \\
        &      & \Delta
    \end{pmatrix}
\end{align*}
where $\nu \geq 1$,
\[\Delta = \begin{cases}
    \emptyset, & \text{if} \ \delta = 0,\\
    (1) \ \text{or}  \ (z), & \text{if} \ \delta=1,\\
    \begin{pmatrix}
        1 & \\
        & -z
    \end{pmatrix}, &\text{if} \ \delta=2,
\end{cases}
\]
and $z$ is a fixed non-square unit in $\mathbb{F}_q$. 
Thus the orthogonal group of degree $2\nu+\delta$ can be denoted by $O_{2\nu+\delta,\Delta}(\mathbb{F}_q)$.
There is an action of $O_{2\nu+\delta,\Delta}(\mathbb{F}_q)$ on $\mathbb{F}_q^{2\nu+\delta}$ defined
as follows
\begin{align*}
    \mathbb{F}_q^{2\nu+\delta}\times O_{2\nu+\delta,\Delta}(\mathbb{F}_q) & \rightarrow \mathbb{F}_q^{2\nu+\delta}\\
    ((x_1,x_2,...,x_{2\nu+\delta}),T) & \mapsto (x_1,x_2,...,x_{2\nu+\delta})T.
\end{align*}
The elements of $O_{2\nu+\delta,\Delta}(\mathbb{F}_q)$ are called {\it orthogonal transformations} with respect to $S_{2\nu + \delta}$. The vector space $\mathbb{F}_q^{2\nu+\delta}$ together with above group action of the orthogonal group $O_{2\nu+\delta,\Delta}(\mathbb{F}_q)$ is called the $2\nu + \delta$-dimensional {\it orthogonal space} over $\mathbb{F}_q$ with respect to $S$. 

Now, let $P$ be an $m$-dimensional vector subspace of $\mathbb{F}_q^{2\nu+\delta}$. By abuse of notation, we also use $P$ to denote a matrix representation of this subspace; that is, $P$ is an $m\times(2\nu+\delta)$ matrix of rank $m$ whose rows form a basis of the subspace $P$.
Then $PS_{2\nu + \delta, \Delta}P^T$ is an $m\times m$ symmetric matrix cogredient to one of following forms
    \begin{align*}
        M(m,2s,s) &=\begin{pmatrix}
            0 & I_s & \\
            I_s & 0 & \\
            &  &  0_{m-2s}
        \end{pmatrix},\\
        M(m,2s+1,s,1) &=\begin{pmatrix}
            0 & I_s & & \\
            I_s & 0 & & \\
            &  &  1 &\\
            &  &  & 0_{m-2s-1}
        \end{pmatrix},\\
        M(m,2s+1,s,z) &=\begin{pmatrix}
            0 & I_s & & \\
            I_s & 0 & & \\
            &  &  z &\\
            &  &  & 0_{m-2s-1}
        \end{pmatrix},\\
        M(m,2s+2,s) &=\begin{pmatrix}
            0 & I_s & & & \\
            I_s & 0 & & & \\
            &  &  1 & &\\
            & & & -z & \\
            &  &  & & 0_{m-2s-2}
        \end{pmatrix}.
    \end{align*}
We write $M(m,2s+\gamma,s,\Gamma)$ for any one of these four normal forms, where $s \geq 1$, $\gamma = 0,1, \ \text{or} \ 2$, and $\Gamma$ represents the definite part in these normal forms. In particular, 
\begin{align*}
    \Gamma =
      \begin{cases}
    \emptyset, & \text{if} \ \gamma = 0,\\
    (1) \ \text{or}  \ (z), & \text{if} \ \gamma = 1,\\
    \begin{pmatrix}
        1 & \\
        & -z
    \end{pmatrix}, &\text{if} \ \gamma = 2,
      \end{cases}
\end{align*}
and may be omitted.
The subspace $P$ is said to be {\it of type} $(m, 2s + \gamma, s, \Gamma)$ if $PS_{2\nu + \delta}P^T$ is cogredient to  $M(m,2s+\gamma,s,\Gamma)$. Subspaces of type $(m, 2s, s), (m, 2s + 1, s, 1), (m, 2s + 1, s, z),$ and $(m, 2s + 2, s)$ are also called {\it subspace of the hyperbolic type, the square type, the non-square type}, and {\it the elliptic type}, respectively. 

In this work, we provide constructions of CS matrices based on an orthogonal space over a finite field of odd characteristic. In fact, we consider one type of subspaces for each construction.

\section{Preliminaries}
Let $q$ be an odd prime power and $\mathbb{F}_q^{2\nu + \delta}$ an orthogonal space. The following ingredients are important for our constructions.

\begin{thm}\cite{Wan}\label{g1}
    The number of subspace of  type $(m,2s,s)$ in  $\mathbb{F}_q^{2\nu + \delta}$, where $2s\leq m\leq \nu+s$ is given by 
    \begin{align*}
        N(m,2s,s;2\nu+\delta,\Delta)
   = q^{2s(\nu +s-m)+\delta s} 
	\frac{\prod\limits_{i=\nu+s-m+1}^{\nu} (q^i-1)(q^{i+\delta-1}+1) }{\prod\limits_{i=1}^{s}(q^i-1) \prod\limits_{i=0}^{s-1}(q^i+1) \prod\limits_{i=1}^{m-2s}(q^i-1)}.
    \end{align*}
\end{thm}

\begin{thm}\cite{Wan}\label{g2}
    The number of subspace of  type $(m,2s+1,s,\Gamma)$ in $\mathbb{F}_q^{2\nu + \delta}$, where 
    $$
    2s+1\leq m\leq \begin{cases}
        \nu+s, & \text{if} \ \delta=0, \ \text{or} \ \delta=1 \ \text{and} \ \Gamma\neq\Delta,\\
        \nu+s+1, & \text{if} \ \delta=2, \ \text{or} \ \delta=1 \ \text{and} \ \Gamma=\Delta,
    \end{cases}
    $$     
is given by 
    \begin{align*}
    N(m,2s+1,s,\Gamma;2\nu+\delta,\Delta)=& q^{2s(\nu +s-m)+s(\delta+1)-(m-2s-1)}\\
	&\cdot \frac{\prod\limits_{i=\nu+s-m+2}^{\nu} (q^i-1)(q^{i+\delta-1}+1) }{\prod\limits_{i=1}^{s}(q^i-1) \prod\limits_{i=0}^{s}(q^i+1) \prod\limits_{i=1}^{m-2s-1}(q^i-1) } n_0(m,2s+1,s,\Gamma;2\nu+\delta,\Delta)
    \end{align*}
    where
    \begin{align*}     
    n_0(m,2s+1,s,\Gamma;2\nu+\delta,\Delta)= \begin{cases}
        q^{\nu-s-1}(q^{\nu+s-m+1}-1), &\text{if} \ \ \delta=0, \\
				q^{\nu-s}(q^{\nu+s-m+1}-1), &\text{if} \ \ \delta=1 \ \ \text{and} \ \ \Gamma \neq \Delta, \\
				q^{\nu-s}(q^{\nu+s-m+1}+1), &\text{if} \ \ \delta=1 \ \ \text{and} \ \ \Gamma = \Delta, \\
				q^{\nu-s}(q^{\nu+s-m+2}+1), &\text{if} \ \ \delta=2.
    \end{cases}
    \end{align*}
\end{thm}

\begin{thm}\cite{Wan}\label{g3}
    The number of subspace of  type $(m,2s+2,s;2\nu+\delta,\Delta)$ in $\mathbb{F}_q^{2\nu + \delta}$, where $2s+2\leq m\leq \nu+s+\Delta$ is given by 
    \begin{align*}
        N(m,2s+2,s;2\nu+\delta,\Delta)= & q^{2s(\nu +s-m)+s(\delta+2)-2(m-2s-2)}\\
	&\cdot\frac{\prod\limits_{i=\nu+s-m+3}^{\nu} (q^i-1)(q^{i+\delta-1}+1) }{\prod\limits_{i=1}^{s}(q^i-1) \prod\limits_{i=0}^{s+1}(q^i+1) \prod\limits_{i=1}^{m-2s-2}(q^i-1) }n_0(m,2s+2,s;2\nu+\delta,\Delta)
    \end{align*}
    where
    \begin{align*}
        n_0(m,2s+2,s;2\nu+\delta,\Delta)= \begin{cases}
        q^{2(\nu-s)-2}(q^{\nu+s-m+1}-1)(q^{\nu+s-m+2}-1), &\text{if} \ \ \delta=0, \\
				q^{2(\nu-s)-1}(q^{\nu+s-m+2}-1)(q^{\nu+s-m+2}+1), &\text{if} \ \ \delta=1, \\
				q^{2(\nu-s)}(q^{\nu+s-m+2}+1)(q^{\nu+s-m+3}+1), &\text{if} \ \ \delta=2.
    \end{cases}
    \end{align*}
\end{thm}

Let $P$ be fixed subspace of type $(m,2s+\gamma,s,\Gamma)$ in $\mathbb{F}_{q}^{2\nu+\delta}$. Denote by $\mathcal{M}(m_1,2s_1+\gamma_1,s_1,\Gamma_1;m,2s+\gamma,s,\Gamma;2\nu+\delta,\Delta)$ the set of subspaces of type $(m_1,2s_1+\gamma_1,s_1,\Gamma_1)$ contained in $P$. Let 
$$
N(m_1,2s_1+\gamma_1,s_1,\Gamma_1;m,2s+\gamma,s,\Gamma;2\nu+\delta,\Delta)  =|\mathcal{M}(m_1,2s_1+\gamma_1,s_1,\Gamma_1;m,2s+\gamma,s,\Gamma;2\nu+\delta,\Delta)|.
$$

\begin{thm}\label{Nempty}\cite{Wan}
	$\mathcal{M}(m_1,2s_1+\gamma_1,s_1,\Gamma_1;m,2s+\gamma,s,\Gamma;2\nu+\delta,\Delta)$ is nonempty if and only if 
    \begin{align*}
        &2s+\gamma \leq m \leq \begin{cases}
			\nu+s+ \min\{ \delta,\gamma \}, &\text{if} \ \gamma \neq \delta , \  \text{or} \ 	\gamma=\delta \ \text{and} \ \Gamma=\Delta,	\\
			\nu+s, &\text{if} \ \gamma=\delta=1 \ \text{and} \ \Gamma_1 \neq \Delta,
			\end{cases}\\
        &2s_1+\gamma_1 \leq m_1 \leq \begin{cases}
			\nu+s_1+ \min\{ \delta,\gamma_1 \}, &\text{if} \ \gamma_1 \neq \delta , \  \text{or} \ 	\gamma_1=\delta \ \text{and} \ \Gamma_1=\Delta,	\\
			\nu+s_1, &\text{if} \ \gamma_1=\delta=1 \ \text{and} \ \Gamma_1 \neq \Delta,\end{cases}\\
        &\text{and}\\
        &2m-2m_1 \geq 
        \begin{cases}
		(2s+\gamma) - (2s_1+\gamma_1) + |\gamma-\gamma_1| \geq 2|\gamma-\gamma_1|, &\text{if} \ \gamma_1 \neq \gamma, \ \text{or} \ \gamma_1=\gamma \ \text{and} \ \Gamma=\Delta, \\
		(2s+\gamma) - (2s_1+\gamma_1) + 2 \geq 4, &\text{if} \ \gamma_1 = \gamma=1 \ \text{and} \  \Gamma_1 \neq \Gamma.
		\end{cases}
    \end{align*}
\end{thm}

\begin{rmk}
    The conditions in Theorem \ref{Nempty} are equivalent to 
    \begin{align*}
        &2s+\gamma \leq m \leq \begin{cases}
	\nu+s+\min\{ \delta,\gamma \}, &\text{if} \ \gamma \neq \delta , \        \text{or} \ 	\gamma=\delta \ \text{and} \ \Gamma=\Delta,	\\
	\nu+s, &\text{if} \ \gamma=\delta=1 \ \text{and} \ \Gamma_1 \neq          \Delta,\end{cases}\\
        &\text{and}\\
        &\min\{ m-2s-\gamma, m_1-2s_1-\gamma_1 \} \geq 
	\begin{cases}
	\max\{ 0,m_1-s-s_1-\min\{ \gamma, \gamma_1\} \}, &\text{if}     \ \gamma_1 \neq \gamma, \ \text{or} \ \gamma_1=\gamma \ \text{and}	\     \Gamma_1=\Gamma, \\
	\max\{ 0, m_1-s-s_1\},  &\text{if} \ \gamma_1=\gamma=1 \         \text{and} \ \Gamma_1 \neq \Gamma. \end{cases} 
    \end{align*}
\end{rmk}

\begin{thm}\label{Nms}\cite{Wan}
If $\mathcal{M}(m_1,2s_1+\gamma_1,s_1,\Gamma_1;m,2s+\gamma,s,\Gamma;2\nu+\delta,\Delta)$ is nonempty, then 
    \begin{align*}
&N(m_1,2s_1+\gamma_1,s_1,\Gamma_1;m,2s+\gamma,s,\Gamma;2\nu+\delta,\Delta)\\
    &=\sum\limits_{k} q^{2s_1(s+s_1-m_1+k)+s_1(\gamma+\gamma_1)-\gamma_1(m_1-2s_1-\gamma_1-k)+(m_1-k)(m-2s-\gamma-k)}\\
    & \ \ \cdot n_0(m_1-k,2s_1+\gamma_1,s_1,\Gamma_1;2s+\gamma,\Gamma)  \frac{ \prod\limits_{i=s+s_1-m_1+\gamma_1+k+1}^{s}(q^i-1)(q^{i+\gamma-1}+1)\prod\limits_{i=m-2s-\gamma-k+1}^{m-2s-\gamma}(q^i-1)}{\prod\limits_{i=1}^{s_1}(q^i-1)\prod\limits_{i=0}^{s_1+\gamma_1-1}(q^i+1)\prod\limits_{i=1}^{m_1-2s_1-\gamma_1-k}(q^i-1)\prod\limits_{i=1}^{k}(q^i-1)}
    \end{align*}
    where the summation range of $k$ is 
    \begin{align*}
        \min\{m-2s-\gamma,m_1-2s_1-\gamma_1\} \geq k \geq 
	\begin{cases}
		\max\{ 0,-s-s_1+m_1+ \min\{\gamma_1,\gamma \},  &\text{if} \ \gamma_1 \neq \gamma, \ \text{or} \ \gamma_1=\gamma \ \text{and} \ \Gamma_1=\Gamma, \\
		\max\{ 0,-s-s_1+m_1 \},  &\text{if} \ \gamma_1=\gamma=1 \ \text{and} \ \Gamma_1 \neq \Gamma,
	\end{cases}
    \end{align*}	    
    and
\begin{enumerate}
	\item[] $n_0(m_1-k,2s_1,s_1;2s+\gamma,\Gamma) = 1$;
	\item[] $n_0(m_1-k,2s_1+1,s_1;2s+\gamma,\Gamma)$
		\subitem =$\begin{cases}
			q^{s-s_1-1}(q^{s+s_1-m_1+k+1}-1), & \text{if} \ \gamma = 0, \\
			q^{s-s_1}(q^{s+s_1-m_1+k+1}+1), & \text{if} \ \gamma = 1 \ \text{and} \ \Gamma_1=\Gamma, \\
			q^{s-s_1}(q^{s+s_1-m_1+k+1}-1), & \text{if} \ \gamma = 1 \ \text{and} \ \Gamma_1 \neq \Gamma, \\
			q^{s-s_1}(q^{s+s_1-m_1+k+2}-1), & \text{if} \ \gamma = 2;
		\end{cases}$
	\item[] $n_0(m_1-k,2s_1+2,s_1;2s+\gamma,\Gamma)$
		\subitem =$\begin{cases}
			q^{2(s-s_1)-2}(q^{s+s_1-m_1+k+1}-1)(q^{s+s_1-m_1+k+2}-1), &\text{if} \ \gamma =0, \\
			q^{2(s-s_1)-1}(q^{s+s_1-m_1+k+2}-1)(q^{s+s_1-m_1+k+2}+1), &\text{if} \ \gamma =1, \\
			q^{2(s-s_1)}(q^{s+s_1-m_1+k+1}+2)(q^{s+s_1-m_1+k+3}+1), &\text{if} \ \gamma = 2.
		\end{cases}$
\end{enumerate}
\end{thm}

\section{CS Matrix Constructions from Orthogonal Spaces}

Let $\mathbb{F}^{2\nu+\delta}_q$ be a $2\nu+\delta$-dimensional orthogonal space over $\mathbb{F}_q$, where $q$ is an odd prime power. Denote set of all subspace type $(m_1, 2s_1+\gamma_1, s_1,\Gamma_1)$ where $2s_1+\gamma_1\leq m_1\leq \nu+s_1+\min\{\delta,\gamma_1\}$ if $\gamma_1\neq\delta$ or $\gamma_1 =\delta$ and $\Gamma=\Delta$ and $2s_1+\gamma_1\leq m_1\leq \nu+s_1$ if $\delta=\gamma_1=1$ and $\Gamma\neq\Delta$, in $\mathbb{F}^{2\nu+\delta}_q$ by $G$, 
\[
G=\{P_1, P_2, ..., P_M\},
\]
where $P_i$'s are different subspaces of type $(m_1,2s_1+\gamma_1,s_1,\Gamma)$ and $M=|G|.$ Denote the set of all subspace of type $(m, 2s+\gamma, s, \Gamma)$, where $2s+\gamma\leq m\leq \nu+s+\min\{\delta,\gamma\}$ if $\gamma\neq\delta$ or $\gamma=\delta$ and $\Gamma=\Delta$ and $2s+\gamma\leq m\leq \nu+s$ if $\delta=\gamma=1$ and $\Gamma\neq\Delta$, in $\mathbb{F}^{2\nu+\delta}_q$ by $H$, 
\[
H=\{Q_1,Q_2,...,Q_N\},
\]
where $Q_j$'s are different subspaces of types $(m, 2s+\gamma, s, \Gamma)$ and $N = |H|$. 

Let $\Phi_0=(a_{ij})$ be an $M \times N$ matrix defined on $\{0,1\}$, where 
\[ 
a_{ij} =\begin{cases}
	1, &\text{ if } P_i\subseteq Q_j, \\
	0, &\text{ else }.
\end{cases} 
\]
From Theorem \ref{g1}, \ref{g2} and \ref{g3}, we have 
\begin{align*}
M &= N(m_1,2s_1+\gamma_1,s_1,\Gamma_1;2\nu+\delta,\Delta), \text{ and } \\
N &= N(m,2s+\gamma,s,\Gamma;2\nu+\delta,\Delta).
\end{align*}

The number of ones in a column of $\Phi_0$ is equivalent to the number of the subspaces of type $(m,2s+\gamma,s,\Gamma)$ that contained in the subspaces of type $(m_1,2s_1+\gamma_1,s_1,\Gamma_1)$. Let $L$ be the number of ones in every column. By Theorem \ref{Nms} we have
\[
L = N(m_1,2s_1+\gamma_1,s_1,\Gamma_1;m,2s+\gamma,s,\Gamma;2\nu+\delta,\Delta).
\]

Let $\Phi_0 =(\alpha_1,\alpha_2,...,\alpha_N)$, where $\alpha_i$'s are column vectors of $\Phi_0.$ Recall that the coherence of $\Phi_0$ is 
\[ 
\mu(\Phi_0)=\max_{1\leq i,j\leq N}\frac{|\langle \alpha_i,\alpha_j \rangle|}{||\alpha_i||_2||\alpha_j||_2}.
\]
Since the matrix $\Phi_0$ is defined on $\{0, 1\}$, $||\alpha_i||^2_2$ equals $L$ and $|\langle \alpha_i,\alpha_j\rangle |$ is the maximum number of subspaces of type $(m_1,2s_1+\gamma_1,s_1,\Gamma_1)$ that can be contained in both two subspaces of type $(m,2s+\gamma,s,\Gamma)$. 
Hence 
\begin{align*}
\max_{1\leq i,j\leq N}|\langle \alpha_i,\alpha_j \rangle| = &N(m_1, 2s_1 + \gamma_1, s_1,\Gamma_1; m-1,2s + \gamma, s, \Gamma)   \text{ or } \\ 
&N(m_1,2s_1+\gamma_1,s_1,\Gamma_1;m-1,2s+\gamma,s-1,\Gamma)
\end{align*}
Let 
$
\Phi=\frac{1}{\sqrt{L}}\Phi_0
$. Then $
\mu=\mu(\Phi)=\mu(\Phi_0)
$
and by Lemma \ref{Bglem}, the matrix $\Phi$ satisfies RIP with order \textit{k}
\[ 
\delta_k\leq (k-1)\mu, 
\]
for any 
\[ 
k<\frac{1}{\mu}+1. 
\]
We also require condition $\delta_{2k} < \sqrt{2} - 1$  to guaranty the exact solution for the basis pursuit problem (\ref{eq:basis-pursuit}).

We propose several distinct constructions of matrices derived from finite orthogonal geometry. In fact, we consider the cases where $P_i$'s and $Q_j$'s lie in the same type family, namely, Hyperbolic type, Square or Non-square type, and Elliptic type.

\section{Parameter Analysis}

 In our construction, we exclude all cases for which the set
\[
\mathcal{M}(m_1,2s_1+\gamma_1,s_1,\Gamma_1;
m,2s+\gamma,s,\Gamma;2\nu+\delta,\Delta)
\]
is empty, since no corresponding incidence matrix can be constructed. This section is devoted to determining the parameters of the resulting matrices, namely the row dimension $M$, the column dimension $N$, and the coherence $\mu$, for each valid case. From the perspective of compressed sensing, a desirable sensing matrix should have as few rows as possible and as many columns as possible while maintaining a sufficiently small coherence. Therefore, for each subspace type, we determine the parameters that minimize $M$ and maximize $N$, and then compute the corresponding coherence $\mu$ to assess the effectiveness of the resulting matrices for compressed sensing. Our analysis is based on logarithmic estimates together with elementary multivariable calculus.

\subsection{When $\gamma = \gamma_1 = 0$ (Hyperbolic type)}

Note that by Theorem \ref{Nempty}, 
\[
\mathcal{M}(m_1,2s_1,s_1;m,2s,s;2\nu+\delta,\Delta)
\]
is nonempty if and only if $\delta=2$, $2s\leq m\leq \nu+s$ and $\min\{m-2s, m_1-s_1\}\geq \max\{0,m_1-s-s_1\}$.
Moreover, if the conditions are satisfied, then
\[
N(m,2s,s;2\nu+2,\Delta)= q^{2s(\nu+s-m)+2s}\frac{\prod\limits_{i=\nu+s-m+1}^{\nu}(q^i-1)(q^{i+1}+1)}{\prod\limits_{i=1}^{s}(q^i-1)\prod\limits_{i=0}^{s-1}(q^i+1)\prod\limits_{i=1}^{m-2s}(q^i-1)},
\]
by Theorem \ref{Nms}. Taking base-$q$ logarithms and using
\[
\log_q(q^i-1)\approx \log_q(q^i+1)\approx i,
\]
we obtain
\begin{align*}
\log_q N(m,2s,s;2\nu+2,\Delta)
\approx&
2s(\nu+s-m)+2s
+\sum_{i=\nu+s-m+1}^{\nu}(2i+1)\\
&-\sum_{i=1}^{s}i
-\sum_{i=0}^{s-1}i
-\sum_{i=1}^{m-2s}i.
\end{align*}
Now,
\[
\sum_{i=\nu+s-m+1}^{\nu}(2i+1)
=
(m-s)(2\nu+s-m+2),
\]
and
\[
\sum_{i=1}^{s}i+\sum_{i=0}^{s-1}i=s^2.
\]
Also,
\[
\sum_{i=1}^{m-2s}i
=
\frac{(m-2s)(m-2s+1)}{2}.
\]
Therefore,
\begin{align*}
\log_q N(m,2s,s;2\nu+2,\Delta)
\approx&
2s(\nu+s-m)+2s
+(m-s)(2\nu+s-m+2)\\
&-s^2
-\frac{(m-2s)(m-2s+1)}{2}\\
=&
-\frac{3}{2}m^2
+\left(2\nu+2s+\frac{3}{2}\right)m
-2s^2+s.
\end{align*}

As $\nu$ is fixed, consider the function
\[
f(x,y)
=
-\frac{3}{2}x^2
+
\left(2\nu+2y+\frac{3}{2}\right)x
-2y^2+y.
\]
We compute
\[
f_x=-3x+2\nu+2y+\frac{3}{2},
\qquad
f_y=2x-4y+1.
\]
Solving $f_x=f_y=0$, we first obtain from $f_y=0$ that
\[
2x-4y+1=0,
\]
so
\[
x=2y-\frac12.
\]
Substituting this into $f_x=0$ gives
\[
-3\left(2y-\frac12\right)
+2\nu+2y+\frac32=0.
\]
Hence
\[
-6y+\frac32+2\nu+2y+\frac32=0,
\]
and therefore
\[
-4y+2\nu+3=0.
\]
Thus
\[
y=\frac{\nu}{2}+\frac34.
\]
Consequently,
\[
x=2\left(\frac{\nu}{2}+\frac34\right)-\frac12
=\nu+1.
\]
Hence the critical point is
\[
\left(\nu+1,\frac{\nu}{2}+\frac34\right).
\]

Next, the Hessian matrix of $f$ is
\[
H_f=
\begin{pmatrix}
-3 & 2\\
2 & -4
\end{pmatrix}.
\]
Since
\[
\det(H_f)=(-3)(-4)-2^2=8>0
\]
and
\[
f_{xx}=-3<0,
\]
the Hessian matrix is negative definite. Therefore, by the second derivative
test, $f$ attains its maximum at
\[
\left(\nu+1,\frac{\nu}{2}+\frac34\right).
\]
Since $m$ and $s$ are integers and the admissibility condition requires
\[
2s\leq m\leq \nu+s,
\]
we choose admissible integer points nearest to this maximum.

If $\nu$ is even, say $\nu=2r$, then the continuous maximum is
\[
\left(2r+1,r+\frac34\right).
\]
The admissible integer choice nearest to this point is
\[
(m,s)=(2r+1,r).
\]
Equivalently, writing $r=s$, this gives
\[
(m,s)=(2s+1,s).
\]

If $\nu$ is odd, say $\nu=2r+1$, then the continuous maximum is
\[
\left(2r+2,r+\frac54\right).
\]
The admissible integer choice nearest to this point is
\[
(m,s)=(2r+2,r+1).
\]
Equivalently, writing $s=r+1$, this gives
\[
(m,s)=(2s,s).
\]

Now, we consider the following two cases.
\begin{enumerate}
\item[(I)] Suppose that $\nu$ is even, say $\nu=2s$. Set
\begin{align*}
M
&=N(1,0,0;2(2s)+2,\Delta)\\
&=
\frac{(q^{2s}-1)(q^{2s+1}+1)}{q-1},\\
N
&=N(2s+1,2s,s;2(2s)+2,\Delta)\\
&=
q^{2s^2}
\frac{
\prod\limits_{i=s}^{2s}(q^i-1)(q^{i+1}+1)}
{
(q-1)\prod\limits_{i=1}^{s}(q^i-1)
\prod\limits_{i=0}^{s-1}(q^i+1)},\\
L
&=N(1,0,0;2s+1,2s,s;2(2s)+2,\Delta)\\
&=
\frac{(q^s-1)(q^{s-1}+1)
\left(q(q^2-1)+(q^{s+1}-1)(q^s+1)\right)}
{2(q-1)^2}.
\end{align*}
Moreover,
\[
\max_{1\leq i\neq j\leq N}
|\langle \alpha_i,\alpha_j\rangle|
=
N(1,0,0;2s,2s,s;2(2s)+2,\Delta)
=
\frac{(q^s-1)(q^{s-1}+1)}{2(q-1)^2}.
\]
Since each column has squared Euclidean norm $L$, the coherence is
\begin{align*}
\mu
&=
\max_{1\leq i\neq j\leq N}
\frac{|\langle \alpha_i,\alpha_j\rangle|}
{\|\alpha_i\|_2\|\alpha_j\|_2}\\
&=
\frac{1}
{q(q^2-1)+(q^{s+1}-1)(q^s+1)}.
\end{align*}
Thus, asymptotically,
\[
\mu
\approx
\frac{1}{q^{2s+1}+q^3}.
\]
Moreover,
\[
M\approx q^{4s}
\qquad\text{and}\qquad
N\approx q^{4s^2+4s}.
\]

To obtain an RIP constant less than $\sqrt{2}-1$, it is sufficient to impose
\[
k<
\frac{\sqrt{2}-1}{2}
\left(q(q^2-1)+(q^{s+1}-1)(q^s+1)\right)
+\frac12.
\]
Indeed, under this condition and by Lemma~\ref{Bglem}, we have
\[
\delta_{2k}\leq (2k-1)\mu<\sqrt{2}-1.
\]
Since
\[
q(q^2-1)+(q^{s+1}-1)(q^s+1)
\approx q^{2s+1}+q^3,
\]
the condition may be written asymptotically as
\[
k<
\frac{\sqrt{2}-1}{2}
\left(q^{2s+1}+q^3\right)
+\frac12.
\]

\item[(II)] Suppose that $\nu$ is odd, say $\nu=2s-1$. Set
\begin{align*}
M
&=N(1,0,0;2(2s-1)+2,\Delta)\\
&=
\frac{(q^{2s-1}-1)(q^{2s}+1)}{q-1},\\
N
&=N(2s,2s,s;2(2s-1)+2,\Delta)\\
&=
q^{2s^2}
\frac{
\prod\limits_{i=s}^{2s-1}(q^i-1)(q^{i+1}+1)}
{
\prod\limits_{i=1}^{s}(q^i-1)
\prod\limits_{i=0}^{s-1}(q^i+1)},\\
L
&=N(1,0,0;2s,2s,s;2(2s-1)+2,\Delta)\\
&=
\frac{(q^s-1)(q^{s-1}+1)}{q-1}.
\end{align*}
Moreover,
\[
\max_{1\leq i\neq j\leq N}
|\langle \alpha_i,\alpha_j\rangle|
=
N(1,0,0;2s-1,2s-2,s-1;2(2s-1)+2,\Delta)
=
(q^{s-1}-1)(q^{s-2}+1).
\]
Hence
\begin{align*}
\mu
&=
\max_{1\leq i\neq j\leq N}
\frac{|\langle \alpha_i,\alpha_j\rangle|}
{\|\alpha_i\|_2\|\alpha_j\|_2}\\
&=
\frac{(q^{s-1}-1)(q^{s-2}+1)}
{\dfrac{(q^s-1)(q^{s-1}+1)}{q-1}}\\
&=
\frac{(q-1)(q^{s-1}-1)(q^{s-2}+1)}
{(q^s-1)(q^{s-1}+1)}.
\end{align*}
Therefore, asymptotically,
\[
\mu
\approx
\frac{q\cdot q^{s-1}\cdot q^{s-2}}
{q^s\cdot q^{s-1}}
=
\frac{1}{q}.
\]
Moreover,
\[
M\approx q^{4s-2}
\qquad\text{and}\qquad
N\approx q^{4s^2}.
\]

To obtain an RIP constant less than $\sqrt{2}-1$, it is sufficient to impose
\[
k<
\frac{\sqrt{2}-1}{2}
\frac{(q^s-1)(q^{s-1}+1)}
{(q-1)(q^{s-1}-1)(q^{s-2}+1)}
+\frac12.
\]
Asymptotically, since $\mu\approx q^{-1}$, this condition becomes
\[
k<
\frac{\sqrt{2}-1}{2}q+\frac12.
\]
\end{enumerate}

\bigskip

    The constructions for the remaining cases are analogous. We summarize the corresponding values of $M$, $N$, and $\mu$ as follows.

\subsection{The case $\gamma=\gamma_1=1$ and $\Gamma=\Gamma_1$ (Square or non-square type)}

Recall from Theorem~\ref{Nempty} that
\[
\mathcal{M}(m_1,2s_1+1,s_1,\Gamma_1;
m,2s+1,s,\Gamma;2\nu+\delta,\Delta)
\]
is nonempty if and only if either
\[
\delta=0,\qquad 2s+1\leq m\leq \nu+s,\qquad
m-2s-1\geq \max\{0,m_1-s-s_1-1\},
\]
or
\[
\delta=1,\qquad 2s+1\leq m\leq \nu+s,\qquad
m-2s-1\geq \max\{0,m_1-s-s_1-1\}.
\]

\subsubsection{The case $\delta=0$}

By Theorem~\ref{Nms}, we have
\begin{align*}
N(m,2s+1,s,\Gamma;2\nu,\Delta)
=&q^{2s(\nu+s-m)+s-(m-2s-1)}  \\
&\cdot
\frac{
\prod\limits_{i=\nu+s-m+2}^{\nu}(q^i-1)(q^{i-1}+1)}
{
\prod\limits_{i=1}^{s}(q^i-1)
\prod\limits_{i=0}^{s}(q^i+1)
\prod\limits_{i=1}^{m-2s-1}(q^i-1)}
q^{\nu-s-1}(q^{\nu+s-m+1}-1).
\end{align*}
For asymptotic estimates, we use
\[
\log_q(q^i-1)\approx \log_q(q^i+1)\approx i.
\]
Thus
\begin{align*}
\log_q N
\approx&
2s(\nu+s-m)+s-(m-2s-1)
+\sum\limits_{i=\nu+s-m+2}^{\nu}(2i-1)\\
&-\sum\limits_{i=1}^{s}i
-\sum\limits_{i=0}^{s}i
-\sum\limits_{i=1}^{m-2s-1}i
+(\nu-s-1)+(\nu+s-m+1)\\
=&-\frac{3}{2}m^2+
\left(2\nu+2s+\frac{1}{2}\right)m
-2s^2-s.
\end{align*}

As $\nu$ is fixed, consider the function
\[
f(x,y)=
-\frac{3}{2}x^2+
\left(2\nu+2y+\frac{1}{2}\right)x
-2y^2-y.
\]
We compute
\[
f_x=-3x+2\nu+2y+\frac12,
\qquad
f_y=2x-4y-1.
\]
Solving $f_x=f_y=0$, from $f_y=0$ we get
\[
2x-4y-1=0,
\]
so
\[
x=2y+\frac12.
\]
Substituting this into $f_x=0$ gives
\[
-3\left(2y+\frac12\right)+2\nu+2y+\frac12=0.
\]
Hence
\[
-6y-\frac32+2\nu+2y+\frac12=0,
\]
and so
\[
-4y+2\nu-1=0.
\]
Thus
\[
y=\frac{\nu}{2}-\frac14.
\]
Consequently,
\[
x=2\left(\frac{\nu}{2}-\frac14\right)+\frac12=\nu.
\]
Therefore the critical point is
\[
\left(\nu,\frac{\nu}{2}-\frac14\right).
\]

The Hessian matrix is
\[
H_f=
\begin{pmatrix}
-3 & 2\\
2 & -4
\end{pmatrix}.
\]
Since
\[
\det(H_f)=(-3)(-4)-2^2=8>0
\]
and
\[
f_{xx}=-3<0,
\]
the Hessian matrix is negative definite. Hence, by the second derivative test,
$f$ attains its maximum at
\[
\left(\nu,\frac{\nu}{2}-\frac14\right).
\]

Since $m$ and $s$ are integers and the admissibility condition requires
\[
2s+1\leq m\leq \nu+s,
\]
we choose admissible integer points nearest to the continuous maximum.

If $\nu$ is even, say $\nu=2r$, then the continuous maximum is
\[
\left(2r,r-\frac14\right).
\]
The point $(2r,r)$ is not admissible, since it would require
\[
2r+1\leq 2r.
\]
Hence the nearest admissible choice is
\[
(m,s)=(2r,r-1).
\]
Equivalently, writing $r=s+1$, we may write this choice as
\[
(m,s)=(2s+2,s).
\]

If $\nu$ is odd, say $\nu=2r+1$, then the continuous maximum is
\[
\left(2r+1,r+\frac14\right).
\]
The nearest admissible integer choice is
\[
(m,s)=(2r+1,r).
\]
Equivalently, writing $r=s$, we obtain
\[
(m,s)=(2s+1,s).
\]

To minimize the row dimension $M$, we choose the smallest admissible row
parameters. Since the row type is $(m_1,2s_1+1,s_1,\Gamma_1)$, the smallest
admissible choice is
\[
m_1=1,\qquad s_1=0.
\]
Thus $M$ is minimized by taking
\[
M=N(1,1,0,\Gamma_1;2\nu,\Delta).
\]

\begin{enumerate}
\item[(I)] Suppose that $\nu$ is even, say $\nu=2s+2$. 
Since $m_1\geq 1$, the minimum value of $M$ is obtained by choosing
\[
m_1=1,\qquad s_1=0.
\]
For the column parameter, the maximizing admissible choice is
\[
(m,s)=(2s+2,s).
\]
Set
\begin{align*}
M
&=N(1,1,0,\Gamma_1;2(2s+2),\Delta)\\
&=
\frac{q^{2s+1}(q^{2s+2}-1)}{2},\\
N
&=N(2s+2,2s+1,s,\Gamma;2(2s+2),\Delta) \\
&=
q^{2s^2+s-1}
\frac{
\prod\limits_{i=s+2}^{2s+2}(q^i-1)(q^{i-1}+1)}
{
(q-1)\prod\limits_{i=1}^{s}(q^i-1)
\prod\limits_{i=0}^{s}(q^i+1)}
q^{s+1}(q^{s+1}-1),\\
L
&=
N(1,1,0,\Gamma_1;
2s+2,2s+1,s,\Gamma;2(2s+2),\Delta).
\end{align*}
Moreover,
\[
\max_{1\leq i\neq j\leq N}
|\langle \alpha_i,\alpha_j\rangle|
=
N(1,1,0,\Gamma_1;
2s+1,2s+1,s,\Gamma;2(2s+2),\Delta).
\]
Hence
\[
\mu
=
\frac{
N(1,1,0,\Gamma_1;
2s+1,2s+1,s,\Gamma;2(2s+2),\Delta)}
{
N(1,1,0,\Gamma_1;
2s+2,2s+1,s,\Gamma;2(2s+2),\Delta)}
.
\]
Asymptotically, we have
\[
L\approx q^{2s+1},
\qquad
\max_{1\leq i\neq j\leq N}
|\langle \alpha_i,\alpha_j\rangle|
\approx q^{2s}.
\]
Therefore,
\[
\mu\approx \frac{q^{2s}}{q^{2s+1}}=\frac{1}{q}.
\]
Moreover,
\[
M\approx \frac{q^{4s+3}}{2}
\qquad\text{and}\qquad
N\approx q^{4s^2+8s+3}.
\]

\item[(II)] Suppose that $\nu$ is odd, say $\nu=2s+1$. 
Since $m_1\geq 1$, the minimum value of $M$ is obtained by choosing
\[
m_1=1,\qquad s_1=0.
\]
For the column parameter, the maximizing admissible choice is
\[
(m,s)=(2s+1,s).
\]
Set
\begin{align*}
M
&=N(1,1,0,\Gamma_1;2(2s+1),\Delta)\\
&=
\frac{q^{2s}(q^{2s+1}-1)}{2},\\
N
&=N(2s+1,2s+1,s,\Gamma;2(2s+1),\Delta)\\
&=
q^{2s^2+s}
\frac{
\prod\limits_{i=s+2}^{2s+1}(q^i-1)(q^{i-1}+1)}
{
\prod\limits_{i=1}^{s}(q^i-1)
\prod\limits_{i=0}^{s}(q^i+1)}
q^s(q^{s+1}-1),\\
L
&=
N(1,1,0,\Gamma_1;
2s+1,2s+1,s,\Gamma;2(2s+1),\Delta).
\end{align*}
Moreover,
\[
\max_{1\leq i\neq j\leq N}
|\langle \alpha_i,\alpha_j\rangle|
=
N(1,1,0,\Gamma_1;
2s,2s,s,\Gamma;2(2s+1),\Delta).
\]
Hence
\[
\mu
=
\frac{
N(1,1,0,\Gamma_1;
2s,2s,s,\Gamma;2(2s+1),\Delta)}
{
N(1,1,0,\Gamma_1;
2s+1,2s+1,s,\Gamma;2(2s+1),\Delta)}
.
\]
Asymptotically, we have
\[
L\approx q^{2s},
\qquad
\max_{1\leq i\neq j\leq N}
|\langle \alpha_i,\alpha_j\rangle|
\approx q^{2s-1}.
\]
Therefore,
\[
\mu\approx \frac{q^{2s-1}}{q^{2s}}=\frac{1}{q}.
\]
Moreover,
\[
M\approx \frac{q^{4s+1}}{2}
\qquad\text{and}\qquad
N\approx q^{4s^2+4s+1}.
\]
\end{enumerate}

\subsubsection{The case $\delta=1$}

By Theorem~\ref{Nms}, if $\Gamma\neq\Delta$, then
\begin{align*}
N(m,2s+1,s,\Gamma;2\nu+1,\Delta)
=& q^{2s(\nu+s-m)+2s-(m-2s-1)}\\
&\cdot
\frac{
\prod\limits_{i=\nu+s-m+2}^{\nu}(q^i-1)(q^i+1)}
{
\prod\limits_{i=1}^{s}(q^i-1)
\prod\limits_{i=0}^{s}(q^i+1)
\prod\limits_{i=1}^{m-2s-1}(q^i-1)}
q^{\nu-s}(q^{\nu+s-m+1}-1).
\end{align*}
If $\Gamma=\Delta$, then the final factor $(q^{\nu+s-m+1}-1)$ is replaced by
$(q^{\nu+s-m+1}+1)$. Thus the two cases have the same leading asymptotic
behavior.

Using
\[
\log_q(q^i-1)\approx \log_q(q^i+1)\approx i,
\]
we obtain
\begin{align*}
\log_q N
\approx&
2s(\nu+s-m)+2s-(m-2s-1)
+\sum\limits_{i=\nu+s-m+2}^{\nu}2i\\
&-\sum\limits_{i=1}^{s}i
-\sum\limits_{i=0}^{s}i
-\sum\limits_{i=1}^{m-2s-1}i
+(\nu-s)+(\nu+s-m+1).
\end{align*}
Now,
\[
\sum\limits_{i=\nu+s-m+2}^{\nu}2i
=
(m-s-1)(2\nu+s-m+2),
\]
and
\[
\sum_{i=1}^{s}i+\sum_{i=0}^{s}i=s(s+1).
\]
Also,
\[
\sum_{i=1}^{m-2s-1}i
=
\frac{(m-2s-1)(m-2s)}{2}.
\]
Hence
\begin{align*}
\log_q N
\approx&
-\frac{3}{2}m^2+
\left(2\nu+2s+\frac{3}{2}\right)m
-2s^2-s.
\end{align*}

As $\nu$ is fixed, consider the function
\[
f(x,y)=
-\frac{3}{2}x^2+
\left(2\nu+2y+\frac{3}{2}\right)x
-2y^2-y.
\]
Then
\[
f_x=-3x+2\nu+2y+\frac{3}{2},
\qquad
f_y=2x-4y-1.
\]
Solving $f_x=f_y=0$ gives
\[
x=\nu+\frac{1}{2},
\qquad
y=\frac{\nu}{2}.
\]
Moreover,
\[
H_f=
\begin{pmatrix}
-3 & 2\\
2 & -4
\end{pmatrix},
\]
so
\[
\det(H_f)=8>0
\qquad\text{and}\qquad
f_{xx}=-3<0.
\]
Thus $f$ attains its maximum at
\[
\left(\nu+\frac12,\frac{\nu}{2}\right).
\]

Since $m$ and $s$ are integers and the admissibility condition requires
\[
2s+1\leq m,
\]
we choose admissible integer points nearest to this maximum. If $\nu$ is even,
say $\nu=2s$, then we choose
\[
(m,s)=(2s+1,s).
\]
If $\nu$ is odd, say $\nu=2s+1$, then the maximizing admissible choice is
\[
(m,s)=(2s+1,s).
\]
Although $(2s+2,s)$ is also admissible in some cases, it gives a smaller
leading exponent.

To minimize the row dimension $M$, we choose the smallest admissible row
parameters:
\[
m_1=1,\qquad s_1=0.
\]

\begin{enumerate}
\item[(I)] Suppose that $\nu$ is even, say $\nu=2s$, and $\Gamma\neq\Delta$.
Set
\begin{align*}
M
&=N(1,1,0,\Gamma_1;2(2s)+1,\Delta)
 =
\frac{q^{4s}-q^{2s}}{2},\\
N
&=N(2s+1,2s+1,s,\Gamma;2(2s)+1,\Delta)\\
&=
q^{2s^2}
\frac{
\prod\limits_{i=s+1}^{2s}(q^i-1)(q^i+1)}
{
\prod\limits_{i=1}^{s}(q^i-1)
\prod\limits_{i=0}^{s}(q^i+1)}
q^s(q^s-1).
\end{align*}
Asymptotically,
\[
M\approx \frac{q^{4s}}{2},
\qquad
N\approx q^{4s^2+2s},
\qquad
\mu\approx \frac{1}{q}.
\]

\item[(II)] Suppose that $\nu$ is even, say $\nu=2s$, and $\Gamma=\Delta$.
Set
\begin{align*}
M
&=N(1,1,0,\Gamma_1;2(2s)+1,\Delta)
 =
\frac{q^{4s}+q^{2s}}{2},\\
N
&=N(2s+1,2s+1,s,\Gamma;2(2s)+1,\Delta)\\
&=
q^{2s^2}
\frac{
\prod\limits_{i=s+1}^{2s}(q^i-1)(q^i+1)}
{
\prod\limits_{i=1}^{s}(q^i-1)
\prod\limits_{i=0}^{s}(q^i+1)}
q^s(q^s+1).
\end{align*}
Asymptotically,
\[
M\approx \frac{q^{4s}}{2},
\qquad
N\approx q^{4s^2+2s},
\qquad
\mu\approx \frac{1}{q}.
\]

\item[(III)] Suppose that $\nu$ is odd, say $\nu=2s+1$, and $\Gamma\neq\Delta$.
Set
\begin{align*}
M
&=N(1,1,0,\Gamma_1;2(2s+1)+1,\Delta)
 =
\frac{q^{4s+2}-q^{2s+1}}{2},\\
N
&=N(2s+1,2s+1,s,\Gamma;2(2s+1)+1,\Delta)\\
&=
q^{2s^2+2s}
\frac{
\prod\limits_{i=s+2}^{2s+1}(q^i-1)(q^i+1)}
{
\prod\limits_{i=1}^{s}(q^i-1)
\prod\limits_{i=0}^{s}(q^i+1)}
q^{s+1}(q^{s+1}-1).
\end{align*}
Asymptotically,
\[
M\approx \frac{q^{4s+2}}{2},
\qquad
N\approx q^{4s^2+6s+2},
\qquad
\mu\approx \frac{1}{q}.
\]

\item[(IV)] Suppose that $\nu$ is odd, say $\nu=2s+1$, and $\Gamma=\Delta$.
Set
\begin{align*}
M
&=N(1,1,0,\Gamma_1;2(2s+1)+1,\Delta)
 =
\frac{q^{4s+2}+q^{2s+1}}{2},\\
N
&=N(2s+1,2s+1,s,\Gamma;2(2s+1)+1,\Delta)\\
&=
q^{2s^2+2s}
\frac{
\prod\limits_{i=s+2}^{2s+1}(q^i-1)(q^i+1)}
{
\prod\limits_{i=1}^{s}(q^i-1)
\prod\limits_{i=0}^{s}(q^i+1)}
q^{s+1}(q^{s+1}+1).
\end{align*}
Asymptotically,
\[
M\approx \frac{q^{4s+2}}{2},
\qquad
N\approx q^{4s^2+6s+2},
\qquad
\mu\approx \frac{1}{q}.
\]
\end{enumerate}

For all subcases in this case (Square or non-square type) for both $\delta = 0$ and $\delta = 1$, the coherence satisfies
\[
\mu\approx \frac{1}{q}.
\]
Therefore, to obtain an RIP constant less than $\sqrt{2}-1$, it is asymptotically sufficient to impose
\[
k<\frac{\sqrt{2}-1}{2}q+\frac{1}{2}.
\]

\subsection{When $\gamma=2$ (Elliptic type)}

By Theorem~\ref{Nempty},
\[
\mathcal{M}(m_1,2s_1+2,s_1;
m,2s+2,s;2\nu+\delta,\Delta)
\]
is nonempty if and only if
\[
\delta=2,\qquad 2s+2\leq m\leq \nu+s+2,
\]
and
\[
\min\{m-2s-2,m_1-2s_1-2\}
\geq
\max\{0,m_1-s-s_1-2\}.
\]
Moreover, if these conditions are satisfied, then by Theorem~\ref{Nms},
\begin{align*}
N(m,2s+2,s;2\nu+2,\Delta)
=&q^{2s(\nu+s-m)+4s-2(m-2s-2)} \\
&\cdot
\frac{
\prod\limits_{i=\nu+s-m+3}^{\nu}(q^i-1)(q^{i+1}+1)}
{
\prod\limits_{i=1}^{s}(q^i-1)
\prod\limits_{i=0}^{s+1}(q^i+1)
\prod\limits_{i=1}^{m-2s-2}(q^i-1)}
\\
&\cdot
q^{2(\nu-s)}
(q^{\nu+s-m+2}+1)(q^{\nu+s-m+3}+1).
\end{align*}

Taking base-$q$ logarithms and using
\[
\log_q(q^i-1)\approx \log_q(q^i+1)\approx i,
\]
we obtain
\begin{align*}
\log_q N
\approx&
2s(\nu+s-m)+4s-2(m-2s-2)
+\sum\limits_{i=\nu+s-m+3}^{\nu}(2i+1)\\
&-\sum\limits_{i=1}^{s}i
-\sum\limits_{i=0}^{s+1}i
-\sum\limits_{i=1}^{m-2s-2}i\\
&+2(\nu-s)
+(\nu+s-m+2)
+(\nu+s-m+3)\\
=&-\frac{3}{2}m^2
+\left(2\nu+2s+\frac{7}{2}\right)m
-2s^2-3s-1.
\end{align*}

Let
\[
f(x,y)=
-\frac{3}{2}x^2
+\left(2\nu+2y+\frac{7}{2}\right)x
-2y^2-3y-1.
\]
Then
\[
f_x=-3x+2\nu+2y+\frac{7}{2},
\qquad
f_y=2x-4y-3.
\]
Solving $f_x=f_y=0$ gives
\[
x=\nu+1,
\qquad
y=\frac{\nu}{2}-\frac14.
\]
Moreover,
\[
H_f=
\begin{pmatrix}
-3 & 2\\
2 & -4
\end{pmatrix},
\]
so
\[
\det(H_f)=8>0
\qquad\text{and}\qquad
f_{xx}=-3<0.
\]
Hence, by the second derivative test, $f$ attains its maximum at
\[
\left(\nu+1,\frac{\nu}{2}-\frac14\right).
\]

Since $m$ and $s$ are integers and the admissibility condition requires
\[
2s+2\leq m,
\]
we choose the nearest admissible integer points to this maximum. If $\nu$ is even, say $\nu=2s+2$, then we may choose the admissible maximizing point
\[
(m,s)=(2s+3,s).
\]
If $\nu$ is odd, say $\nu=2s+1$, then we choose
\[
(m,s)=(2s+2,s).
\]

To find the minimum of $N$, for fixed $s$ consider
\[
g(x)=
-\frac{3}{2}x^2
+\left(2\nu+2s+\frac{7}{2}\right)x
-2s^2-3s-1.
\]
Since $g$ is a downward-opening parabola, its minimum over the admissible interval is attained at an endpoint. Taking the smallest admissible parameters gives
\[
m=2,\qquad s=0.
\]
Thus the minimum is attained at
\[
N(2,2,0;2\nu+2,\Delta).
\]

\begin{enumerate}
  
\item[(I)] Suppose that $\nu$ is even, say $\nu=2s+2$.
For the row parameter, the minimum value of $M$ is obtained by choosing
\[
m_1=2,\qquad s_1=0.
\]
For the column parameter, we choose the admissible maximizing value
\[
(m,s)=(2s+3,s).
\]
Set
\begin{align*}
M
&=N(2,2,0;2(2s+2)+2,\Delta)\\
&=
\frac{
q^{4s+4}(q^{2s+2}+1)(q^{2s+3}+1)}
{2(q+1)},\\
N
&=N(2s+3,2s+2,s;2(2s+2)+2,\Delta)\\
&=
q^{2s^2+2s-2}
\frac{
\prod\limits_{i=s+2}^{2s+2}(q^i-1)(q^{i+1}+1)}
{
(q-1)\prod\limits_{i=1}^{s}(q^i-1)
\prod\limits_{i=0}^{s+1}(q^i+1)}
q^{2s+4}(q^{s+1}+1)(q^{s+2}+1),\\
L
&=
N(2,2,0;
2s+3,2s+2,s;2(2s+2)+2,\Delta).
\end{align*}
Moreover,
\[
\max_{1\leq i\neq j\leq N}
|\langle \alpha_i,\alpha_j\rangle|
=
N(2,2,0;\mathcal I;2(2s+2)+2,\Delta),
\]
where $\mathcal I$ denotes the intersection type giving the largest intersection
among two distinct columns. Hence
\[
\mu
=
\frac{
N(2,2,0;\mathcal I;2(2s+2)+2,\Delta)}
{
N(2,2,0;
2s+3,2s+2,s;2(2s+2)+2,\Delta)}
.
\]

Asymptotically, the column norm satisfies
\[
L\approx q^{4s+2},
\]
and the largest intersection number is of order
\[
\max_{1\leq i\neq j\leq N}
|\langle \alpha_i,\alpha_j\rangle|
\approx q^{4s}.
\]
Therefore,
\[
\mu
\approx
\frac{q^{4s}}{q^{4s+2}}
=
\frac{1}{q^2}.
\]

Moreover,
\[
M\approx \frac{q^{8s+8}}{2},
\text{ and }
N\approx q^{4s^2+12s+8}.
\]

\item[(II)] Suppose that $\nu$ is odd, say $\nu=2s+1$.
For the row parameter, the minimum value of $M$ is obtained by choosing
\[
m_1=2,\qquad s_1=0.
\]
For the column parameter, we choose the admissible maximizing value
\[
(m,s)=(2s+2,s).
\]
Set
\begin{align*}
M
&=N(2,2,0;2(2s+1)+2,\Delta)\\
&=
\frac{
q^{4s+2}(q^{2s+1}+1)(q^{2s+2}+1)}
{2(q+1)},\\
N
&=N(2s+2,2s+2,s;2(2s+1)+2,\Delta)\\
&=
q^{2s^2+4s+2}
\frac{
\prod\limits_{i=s+2}^{2s+1}(q^i-1)(q^{i+1}+1)}
{
\prod\limits_{i=1}^{s}(q^i-1)
\prod\limits_{i=0}^{s+1}(q^i+1)}
(q^{s+1}+1)(q^{s+2}+1)\\
L
&=
N(2,2,0;
2s+2,2s+2,s;2(2s+1)+2,\Delta).
\end{align*}
Moreover,
\[
\max_{1\leq i\neq j\leq N}
|\langle \alpha_i,\alpha_j\rangle|
=
N(2,2,0;\mathcal I;2(2s+1)+2,\Delta),
\]
where $\mathcal I$ denotes the intersection type giving the largest intersection
among two distinct columns. Hence
\[
\mu
=
\frac{
N(2,2,0;\mathcal I;2(2s+1)+2,\Delta)}
{
N(2,2,0;
2s+2,2s+2,s;2(2s+1)+2,\Delta)}
.
\]
Asymptotically, the column norm satisfies
\[
L\approx q^{4s},
\]
and the largest intersection number is of order
\[
\max_{1\leq i\neq j\leq N}
|\langle \alpha_i,\alpha_j\rangle|
\approx q^{4s-2}.
\]
Therefore,
\[
\mu
\approx
\frac{q^{4s-2}}{q^{4s}}
=
\frac{1}{q^2}.
\]

Moreover,
\[
M\approx \frac{q^{8s+4}}{2},
\text{ and },
N\approx q^{4s^2+8s+4}.
\]
\end{enumerate}

{\color{red}

}

 For both subcases (I) - (II), the coherence satisfies
\[
\mu\approx \frac{1}{q^2}.
\]
Therefore, to obtain an RIP constant less than $\sqrt{2}-1$, it is asymptotically sufficient to impose
\[
k<\frac{\sqrt{2}-1}{2}q^2+\frac{1}{2}.
\]
Indeed, under this condition and by Lemma~\ref{Bglem}, we have
\[
\delta_{2k}\leq (2k-1)\mu<\sqrt{2}-1.
\]

In 
Table~\ref{tab:parameters} we collect the essential asymptotic parameters obtained in the
preceding subsections. For each subspace type, the table records the selected
parameters, the resulting row dimension $M$, column dimension $N$, coherence
$\mu$, and the corresponding sufficient sparsity condition for obtaining an RIP
constant less than $\sqrt{2}-1$. This summary also indicates which constructions
provide meaningful sparse recovery guarantees and which ones have limited
compressed sensing performance due to relatively large coherence.

\begin{table}[ht]
\centering
\caption{Asymptotic parameters of the constructed matrices.}
\label{tab:parameters}
\renewcommand{\arraystretch}{1.4}
\resizebox{\textwidth}{!}{%
\begin{tabular}{|c|c|c|c|c|c|}
\hline
Type &
Case &
Rows $M$ &
Columns $N$ &
Coherence $\mu$ &
Sufficient condition on $k$  \\ \hline

Hyperbolic &
$\nu=2s$ &
$\approx q^{4s}$ &
$\approx q^{4s^2+4s}$ &
$\approx (q^{2s + 1} + q^3)^{-1}$ &
$k<\dfrac{\sqrt{2}-1}{2}(q^{2s + 1} + q^3)+\dfrac12$ \\ \hline

Hyperbolic &
$\nu=2s-1$ &
$\approx q^{4s-2}$ &
$\approx q^{4s^2}$ &
$\approx q^{-1}$ &
$k<\dfrac{\sqrt{2}-1}{2}q+\dfrac12$ \\ \hline

Square / Non-square, $\delta=0$ &
$\nu=2s+2$ &
$\approx \frac{q^{4s+3}}{2}$ &
$\approx q^{4s^2+8s+3}$ &
$\approx q^{-1}$ &
$k<\dfrac{\sqrt{2}-1}{2}q+\dfrac12$ \\ \hline

Square / Non-square, $\delta=0$ &
$\nu=2s+1$ &
$\approx \frac{q^{4s+1}}{2}$ &
$\approx q^{4s^2+4s+1}$ &
$\approx q^{-1}$ &
$k<\dfrac{\sqrt{2}-1}{2}q+\dfrac12$ \\ \hline

Square / Non-square, $\delta=1$ &
$\nu=2s,\ \Gamma\neq\Delta$ &
$\approx \frac{q^{4s}}{2}$ &
$\approx q^{4s^2+2s}$ &
$\approx q^{-1}$ &
$k<\dfrac{\sqrt{2}-1}{2}q+\dfrac12$ \\ \hline

Square / Non-square, $\delta=1$ &
$\nu=2s,\ \Gamma=\Delta$ &
$\approx \frac{q^{4s}}{2}$ &
$\approx q^{4s^2+2s}$ &
$\approx q^{-1}$ &
$k<\dfrac{\sqrt{2}-1}{2}q+\dfrac12$ \\ \hline

Square / Non-square, $\delta=1$ &
$\nu=2s-1,\ \Gamma\neq\Delta$ &
$\approx \frac{q^{4s + 2}}{2}$ &
$\approx q^{4s^2+6s+2}$ &
$\approx q^{-1}$ &
$k<\dfrac{\sqrt{2}-1}{2}q+\dfrac12$ \\ \hline

Square / Non-square, $\delta=1$ &
$\nu=2s-1,\ \Gamma=\Delta$ &
$\approx \frac{q^{4s + 2}}{2}$ &
$\approx q^{4s^2+6s+2}$ &
$\approx q^{-1}$ &
$k<\dfrac{\sqrt{2}-1}{2}q+\dfrac12$ \\ \hline

Elliptic &
$\nu=2s$ &
$\approx \frac{q^{8s+8}}{2}$ &
$\approx q^{4s^2+12s+8}$ &
$\approx q^{-2}$ &
$k<\dfrac{\sqrt{2}-1}{2}q^2+\dfrac12$ \\ \hline

Elliptic &
$\nu=2s-1$ &
$\approx \frac{q^{8s+4}}{2}$ &
$\approx q^{4s^2+8s+4}$ &
$\approx q^{-2}$ &
$k<\dfrac{\sqrt{2}-1}{2}q^2+\dfrac12$ \\ \hline
\end{tabular}}
\end{table}

The following example illustrates how the parameter estimates can be used in
practice. Given a prescribed signal dimension and sparsity level, one may first
choose parameters so that the construction produces sufficiently many columns,
and then select the required number of columns. Since passing to a submatrix by
selecting columns does not increase the coherence, the same coherence-based RIP
guarantee remains valid.

\begin{ex}
Suppose that the desired signal dimension is $N=10^6$ and the target sparsity is
$k=100$. Consider the hyperbolic construction with $q=7$ and $s=1$, so that
$\nu=2$. In this case,
\[
M=
\frac{(q^2-1)(q^3+1)}{q-1}
=
\frac{(7^2-1)(7^3+1)}{7-1}
=
2752.
\]
The full construction produces
\begin{align*}
N_0
&=
q^2
\frac{
(q-1)(q^2+1)(q^2-1)(q^3+1)}
{2(q-1)^2} \\
&=
7^2
\frac{
(7-1)(7^2+1)(7^2-1)(7^3+1)}
{2(7-1)^2} \\
&=
3{,}371{,}200
>
10^6
\end{align*}
columns. Hence we may select any $10^6$ columns to obtain a matrix
\[
\Phi\in\mathbb{R}^{2752\times 10^6}.
\]
Since selecting a subset of columns does not increase the coherence, the
coherence of $\Phi$ is at most
\[
\mu
=
\frac{1}{q(q^2-1)+(q^2-1)(q+1)}
=
\frac{1}{720}.
\]
Therefore, for $k=100$,
\[
(2k-1)\mu
=
199\cdot \frac{1}{720}
=
\frac{199}{720}
\approx 0.2764
<
\sqrt{2}-1.
\]
By Lemma~\ref{Bglem},
\[
\delta_{200}\leq (2k-1)\mu<\sqrt{2}-1.
\]
Thus, $\Phi$ satisfies the RIP condition required for exact recovery of
$100$-sparse signals via Basis Pursuit.
\end{ex}

 \section{Numerical Comparison with DeVore's Deterministic Construction}

In this section, we compare the proposed matrix constructions with DeVore's
deterministic construction. The comparison is made in terms of the number of
rows $M$, the number of columns $N$, the coherence $\mu$, and the sparsity level
guaranteed by the coherence-based RIP condition.

DeVore's construction produces a matrix with parameters
\[
M_D=p^{2d},\qquad N_D=p^{r+1},\qquad \mu_D\leq \frac{r}{p}.
\]
Hence, by Lemma~\ref{Bglem}, exact recovery of $k$-sparse signals is guaranteed
whenever
\[
(2k-1)\mu_D<\sqrt{2}-1.
\]
Equivalently,
\[
k<\frac{\sqrt{2}-1}{2}\frac{p^d}{r}+\frac12.
\]

For a meaningful comparison, we choose DeVore parameters $(p^d,r)$ so that the
number of columns $N_D=p^{d(r+1)}$ is comparable to the number of columns $N$ in
our construction. Since $N$ represents the ambient signal dimension, this allows
us to compare the number of required measurements and the corresponding recovery
guarantees for signals of comparable length. 
It is worth noting that the comparison with DeVore's construction depends on
the choice of the parameters $(p^d,r)$. For example, in the hyperbolic case with
$s=2$, our construction has
\[
M\approx q^8,\qquad N\approx q^{24}.
\]
If DeVore's parameters are chosen so that
\[
N_D=p^{d(r+1)}\approx q^{24},
\]
then
\[
p^d \approx q^{24/(r+1)}
\]
and hence
\[
M_D=p^{2d}\approx q^{48/(r+1)}.
\]
Therefore, our construction has fewer rows than DeVore's construction whenever
\[
q^8<q^{48/(r+1)},
\]
or equivalently,
\[
r\leq 4.
\]
For larger values of $r$, DeVore's construction uses fewer rows. Thus, the
comparison should be interpreted as a trade-off depending on the chosen
parameters, rather than as a uniform dominance of one construction over the
other. For this hyperbolic case with $\nu = 2s$, we present the row dimensional comparison between the two matrices from our construction and DeVore's, respectively. Let $\Phi$ be the matrix we construct and the parameters are 
\[
M\approx q^{4s}, N\approx q^{4s^2+4s}, \mu\approx \frac{1}{q^{2s+1}+q^3}.
\]
Let $\Phi_1$ be the matrix under the DeVore's construction with the prime power $q^{4s}$, and $r=s$. Then the parameters are
\[
M_1=q^{8s}, N_1=q^{4s^2+4s}, \mu_1=\frac{s}{q^{4s}}.
\]
Note that $\mu_1$ is less than $\mu$.
By Lemma \ref{Bglem}, 
if a signal $x$ of sparsity \textit{k} satisfies $k\leq\frac{1}{\mu} \leq \frac{1}{\mu_1}$ and $k < \frac{\sqrt{2} - 1}{2}(q^{2s + 1} + q^3) + \frac{1}{2}$, then both $\Phi$ and $\Phi_1$ satisfy RIP and are useful for the basis pursuit problem. We conclude that the column dimension $N = N_1$, but our $M$ is smaller than $M_1$.

\medskip

In Table~\ref{tab:devore-comparison},
we present another example by taking $p^d=q=101$ and $s=2$, and choose $r$ so that $N_D=N$ at the level of
powers of $q$. The value $k_{\max}$ is computed from
\[
k_{\max}
=
\left\lfloor
\frac{\sqrt{2}-1}{2\mu}+\frac12
\right\rfloor .
\]

\begin{table}[ht]
\centering
\caption{Numerical comparison with DeVore's construction for comparable column dimensions.}
\label{tab:devore-comparison}
\renewcommand{\arraystretch}{1.35}
\resizebox{\textwidth}{!}{
\begin{tabular}{|c|c|c|c|c|c|c|c|c|}
\hline
Construction &
$q$ &
$s$ &
$M$ &
$N$ &
$\mu$ &
$k_{\max}$ &
DeVore $(p^d,r)$ &
DeVore $(M_D,N_D,\mu_D,k_D)$ \\ \hline

Hyperbolic, $\nu=2s$ &
$101$ &
$2$ &
$q^8$ &
$q^{24}$ &
$(q^5+q^3)^{-1}$ &
$2{,}176{,}713{,}085$ &
$(101,31)$ &
$(q^2,q^{24},31/101,1)$ \\ \hline

Square/Non-square, $\delta=0$, $\nu=2s+2$ &
$101$ &
$2$ &
$\frac{q^{11}}{2}$ &
$q^{35}$ &
$q^{-1}$ &
$21$ &
$(101,31)$ &
$(q^2,q^{35},31/101,1)$ \\ \hline

Square/Non-square, $\delta=1$, $\nu=2s$, $\Gamma\neq\Delta$&
$101$ &
$2$ &
$\frac{q^8}{2}$ &
$q^{20}$ &
$q^{-1}$ &
$21$ &
$(101,19)$ &
$(q^2,q^{20},19/101,1)$ \\ \hline

Elliptic, $\nu=2s$ &
$101$ &
$2$ &
$\frac{q^{24}}{2}$ &
$q^{38}$ &
$q^{-2}$ &
$2{,}113$ &
$(101,23)$ &
$(q^2,q^{38},23/101,1)$ \\ \hline

\end{tabular}}
\end{table}

This shows a clear
trade-off between the two approaches. For comparable column dimensions,
DeVore's construction requires substantially fewer rows. However, the proposed
constructions may have significantly smaller coherence, which leads to larger
coherence-based sparsity guarantees. Thus, based on these examples, DeVore's construction is more
efficient in terms of the number of measurements, while some of our constructions
provide stronger recovery guarantees from the coherence-based RIP estimate.

\section*{Acknowledgments}
This project is funded by National Research Council of Thailand (NRCT) and King Mongkut's University of Technology Thonburi (N42A680277). The first author acknowledges the financial support provided by the Center of Excellence in Theoretical and Computational Science (TaCS-CoE), KMUTT. The authors thank the referee for several valuable comments and suggestions, which greatly help to improve the exposition of this paper.

\end{document}